# Proto-CIRCUS Tilted-Coil Tokamak-Torsatron Hybrid: Design and Construction

A. W. Clark[1,2], M. Doumet[1], K. C. Hammond[1], Y. Kornbluth[3], D. A. Spong[4], R. Sweeney[1], F. A. Volpe[1,5]

[1] Dept Applied Phys. & Applied Math., Columbia University, New York, NY 10027, USA

[2] Current address: Dept Physics & Nuclear Eng., US Military Academy at West Point, NY 10996, USA

[3] Yeshiva University, New York, NY 10033, USA

[4] Oak Ridge National Laboratory, Oak Ridge, TN 37830, USA

[5] Corresponding author e-mail: fvolpe@columbia.edu

Abstract

We present the field-line modeling, design and construction of a prototype circular-coil tokamak-torsatron hybrid called Proto-CIRCUS. The device has a major radius $R = 16$ cm and minor radius $a < 5$ cm. The six "toroidal field" coils are planar as in a tokamak, but they are tilted. This, combined with induced or driven plasma current, is expected to generate rotational transform, as seen in field-line tracing and equilibrium calculations. The device is expected to operate at lower plasma current than a tokamak of comparable size and magnetic field, which might have interesting implications for disruptions and steady-state operation. Additionally, the toroidal magnetic ripple is less pronounced than in an equivalent tokamak in which the coils are not tilted. The tilted coils are interlocked, resulting in a relatively low aspect ratio, and can be moved, both radially and in tilt angle, between discharges. This capability will be exploited for detailed comparisons between calculations and field-line mapping measurements. Such comparisons will reveal whether this relatively simple concept can generate the expected rotational transform.

*Keywords—Stellarator, electron cyclotron resonance, interlinked coils, rotational transform, tilted coils*

## 1. Introduction

The rotational transform in stellarator and heliotron devices is generated in full or in part by coils, typically of complicated non-planar shapes [1]. However, stellarator configurations can also rely on simpler, planar (often circular) coils. This is for example the case of the heliac configuration [2] and of the Columbia Non-neutral Torus (CNT) [3]. The latter utilizes two interlinked coils with additional coils for vertical stabilization, similar to an earlier proposal by Todd [4]. Similar low aspect-ratio concepts featuring more than two coils were modeled in the past [5]. Note that, to generate rotational transform, it suffices to tilt the coils. Having them interlinked helps to lower the aspect-ratio of the plasma, but is not strictly required [6]. Such a configuration still requires a vertical field for radial force balance and might require an induced or driven plasma current $I_p$, like a tokamak. Hence we refer to this configuration as a tilted-coil tokamak-stellarator hybrid or, more precisely, as a tokamak-torsatron hybrid, as the tilted coil-currents all point in the same directions, when viewed in a poloidal cross-section. The configuration in which coils are energized in alternate directions, similar to a classical stellarator, will be the subject of future study.

Simulations in Sec.2 show that the $I_p$ requirements for this tokamak-torsatron concept are significantly reduced compared with a tokamak; i.e., significantly lower plasma currents are needed to generate comparable profiles of safety factor $q$. The design also offers reduced effective ripple [7] compared with tokamaks [8]. In Sec.3 we describe the design and the construction of a table-top proof-of-concept utilizing six interlocked toroidal field (TF) coils. Sec.4 describes planned physics experiments and engineering improvements.

## 2. Concept

The present experimental work was motivated by the numerical model of an ITER-scale tokamak-stellarator hybrid featuring 18 tilted interlinked coils [8]. Proto-CIRCUS is a simplified, scaled-down proof of concept utilizing 6 tilted coils instead of 18. Fig.1 shows top, side and bird's-eye views of the 6- and 18-coil device, along with plasma equilibria calculated with the VMEC code [10] and color-contoured according to field strength. Profiles of safety factor *q* (Fig.2) and effective ripple (Fig.3) were obtained respectively from the VMEC and NEO code [7], showing that a device with 18 coils tilted by 40° and a plasma current $I_p$=5-8 MA is characterized by a higher rotational transform and smaller effective ripple than a 18-coil tokamak of $I_p$=8 MA.

Fig.2b suggests that the *q*-profile is relatively insensitive to the tilt angle. However, it should be noted that the innermost radial position of the magnetic axis was obtained for a tilt of 40°. Changing the tilt in either way affected the VMEC equilibrium and tended to shift the magnetic axis toward outer radii, closer to the TF coils. For this reason, as the tilt was changed the plasma became more three-dimensional, increasing the ripple (not shown). The tilt variation studies have just recently been initiated and may require different locations and current distributions for the vertical field coils in order to maintain a constant magnetic axis location than what has been tested.

Fig. 3 indicates that as the number of TF coils is increased, the plasma became less three-dimensional, as evidenced by reduced levels of effective ripple.

The TF coil radius was set to 16.0 cm and all other length-scales were adjusted accordingly. The TF coils are typically positioned at a major radius of 10.7 cm, but this position is adjustable by about ±1cm. The tilt of the coils is also adjustable, in the range 34-61° relative to the horizontal. The radial and vertical position of the coils and their dimensions are summarized in Table 1.

A 1 kW magnetron is used for Electron Cyclotron (EC) start-up, heating and current drive at the first harmonic in the Ordinary (O) mode. For this reason, and because the magnetron frequency is 2.45 GHz, the TF coil currents are chosen to generate a magnetic field of 875 Gauss in the plasma center. Maintaining the relative current ratios from the original model resulted in the coil currents listed in Table 1. The vacuum field generated by such coil-currents was calculated and the field-lines traced (Fig.4), by means of the COMSOL finite element software. It was found that, for $I_p$=2.5 kA, the last closed flux surface has an approximate radius of 4.5 cm. This was corroborated by Poincaré plots calculations (Fig.5). These plots also show how the plasma minor radius decreases as the plasma current is reduced to 1.6 and 0.8 kA, while all coil-currents are kept fixed. Note however that there are other possible sets of coil-currents, currently under numerical study, which might lead to a plasma equilibrium of comparable size but reduced $I_p$.

3. DESIGN AND CONSTRUCTION

*3.1 Coils*

All coils in CIRCUS are circular and multi-turn. Each coil is wound with 3.2 mm thick copper wire carrying about 75 A per turn. The number of turns and electric resistance of the TF, QF and VF coils are shown in Table 1. The coils are expected to experience an amenable temperature increase per unit time, $\Delta T/\Delta t$=0.4 K/s.

For simplicity a single power supply energizes all coils, which for this purpose are properly arranged in series and parallel with each other. In particular the TF coils are combined into three series of two, the QF coils are in series with each other, and the VF coils are in parallel. Voltage dividers are used to set the QF and VF coil series to the correct voltage, which is lower than for the TF coils (Table 1).

To form the TF coils, we used chromium-coated stainless steel bicycle rims of 14.75 cm radius and 3 cm width, which when wound, result in an effective coil radius of 16.0 cm. The rims were cut, interlocked, and welded closed. The next step in coil fabrication was to weld a single axle into each rim, which was then cut out, leaving an inboard and outboard axle on each rim. These two axles help support each coil respectively on the central steel column and on an additional structure at outer radius (Sec.3.3). After installation, the coils were coated in Teflon.

The device is steady-state and the initial and final current transients will be slow (2s). A numerical model shows that, as a result, currents induced in the coil-supports will heat them by less than 1 K.

*3.2 Vacuum vessel*

The coils are internal to cylindrical vessel of 60 cm diameter and 50 cm height. The cylindrical tube is made of acrylic: its transparency to microwaves and visible light make it attractive both for heating and diagnostics. The caps to the cylinder are made of 304 stainless steel. In the center of the cylinder is a stainless steel column to provide additional support to the caps and to provide an inboard attachment point for the coils (Fig. 6). A vacuum force of $1.1\times10^5$ N is exerted on the steel caps and a force of $3.8\times10^4$ N is exerted on the side of the acrylic cylinder.

The compressive forces on the cylinder and the deflection forces on the caps were used to determine the required material thicknesses. Acrylic has a compressive strength of 12.4 kN/cm$^2$. Given the compression between the top and bottom of the cylinder, the acrylic cylinder must be 0.05 cm thick. The radial forces, which translate into a compressive force into the circumference of the cylinder, require a thickness of 0.15 cm. A cylinder 1.25 cm thick was chosen to ensure a sufficient safety margin.

The vertical force causes a deflection of the plates. By means of standard stress and strain calculations [9], the deflection behavior of both an acrylic and steel cap can be calculated as functions of their total thickness, as seen in Fig. 7. Based on this information and the need to machine sealable ports onto the caps, 1.7 cm thick 304 stainless steel was chosen for the cap. The actual vacuum chamber deflection is less than 1mm.

The chambers features a total of 6 ports, all on the bottom cap: three for electric feed-throughs, one for a gas feed-through, one for pressure gauges, and one for the vacuum pump.

All seals were accomplished by means of Viton O-rings. The chamber reached a pressure of $2.2\times10^{-5}$ Torr. When pumped down from atmospheric pressure, the chamber starts with a leak rate of $3\times10^{-5}$ Torr/s and after 36 hours, reaches a leak rate of $1.4\times10^{-5}$ Torr/s, consistent with typical acrylic outgassing rates. We know from separate experiments on CNT that this pressure of neutrals is adequate for EC start-up.

Care was taken not to use acetone on acrylic or rubber to check for vacuum leaks, as it can break the acrylic chamber. Methanol is also not advised unless removed with water immediately. Leaks were checked with methanol, followed by immediate water rinses.

*3.3 Coil Supports*

Each coil is attached by two axles to the central stainless steel column and to a 1 cm thick copper bracket located inside the chamber and mounted on the bottom plate (Fig. 6). The brackets and central column also serve to hold the VF and QF coils. The brackets are not attached to the top plate, to ease its removal (Fig. 6).

The coil supports are made of copper, due to its high thermal conductivity. This is to facilitate the conduction of the 24 kJ of heat deposited in each coil during a typical discharge. Heat is conducted to the top and bottom plates and eventually removed by water cooling external to the vessel.

A model of the coil cooling time was developed based on the following assumptions: 1) all heat conducts through the brackets, 2) there is perfect thermal contact between different surfaces, and 3) the top and bottom plates are cooled to $10^\circ$ C. In a 30 s long discharge, long enough for Langmuir probes and flux-surface diagnostics to be manually scanned (Sec.4.1), the coils warm up by $12^\circ$ C. Note that a new discharge can start before complete recovery of the original $10^\circ$ C temperature, but there will be a temperature pile-up effect. For example, three 30 s discharges can be run in rapid sequence, interleaved with 4-8 minutes cooling times, but the coils will reach temperatures of about $42^\circ$ C, after which it will take them approximately 23 minutes to cool to $10^\circ$ C again. These estimates can be easily adapted to shorter discharges: since the cooling rate is an exponential function of the relative temperature between the coils and the steel plate, shorter plasma discharges result in exponentially shorter cooling times.

The radial position of the TF coils is continuously and independently adjustable by means of slots in the supporting brackets, where they attach to the steel bottom plate. The tilt of the TF coils is also continuously adjustable, between 34 and $61^\circ$ relative to the horizontal, thanks to curved slots where the TF coils attach to the bracket (Fig. 8). Because coils and supports are internal to the vessel, radial and tilt adjustments of the coils require breaking the vacuum.

The coil-support design was evaluated for Lorentz forces. The magnetic forces in Proto-CIRCUS apply a torque on the TF coils, tending to rotate the coils about their mid-plane axis. The calculated torque -less than 1 Nm- was experimentally confirmed to put negligible strain on a prototype bracket.

*3.4 Assembly*

Fig.9a shows the acrylic chamber, central column and top and bottom plates assembled during the successful vacuum test described in Sec.2.3. Note the holes in the central column. For the final assembly of the device the interlinked tilted coils were mounted with their studs in such holes (Fig.9b-c). This assembly and the coil support structures were then placed on the bottom plate (on which the lower QF and lower VF coil had previously been installed as in Fig.6) and were fixed to each other. Finally the upper QF and VF coils were also installed (Fig.6) and the chamber closed with the top plate. The final assembly is shown in Fig.10.

*3.5 Magnetron*

A 1kW 2.45 GHz magnetron will be used to ionize the gas and for Electron Cyclotron Current Drive (ECCD). The diverging microwave beam will be first collimated by means of a paraboloidal mirror and then focused by means of a second, steerable paraboloidal mirror. This will allow for heating and current drive in a specific, adjustable location. The experiment will be surrounded by a metal mesh to shield from stray microwave radiation.

## 4. FUTURE WORK AND EXPERIMENTAL PLANS

Proto-CIRCUS is a proof of principle device allowing for modular upgrades and installation of multiple diagnostics.

### *4.1 Diagnostics*

The acrylic vessel provides vast access to optical and X-ray diagnostics. Ports and feed-throughs allow the installation of 6 diagnostics, expandable to a total of 30. The first diagnostics to be installed will be thermocouples, a Langmuir probe, a Rogowski coil for plasma current measurement, and an electron gun and fluorescent screen for flux surface mapping.

### *4.2 Experimentation*

The flux surfaces will be mapped and plasma current measured, for various radial locations and tilt angles of the TF coils. Detailed comparisons with field-line tracing results and VMEC [10] equilibria will allow confirming whether the relatively simple Proto-CIRCUS concept can generate the expected rotational transform. Experimental and numerical scans will identify the most "stellarator-like" configuration, which can be run at the lowest $I_p$. Note that the electron beam mapping of flux surfaces will be carried out in vacuum, in the absence of plasma. The plasma current will thus be simulated by adding a coil at the location of the magnetic axis. A similar technique was used with success for the calibration of magnetic probes in the HBT-EP tokamak [11].

In addition or alternative to ECCD and possibly to a central solenoid, we are also considering starting and sustaining the plasma by means of a plasma gun, as in the Proto-CLEO stellarator [12].

Another area of study is plasma formation by electron cyclotron start-up. The transparent acrylic vessel offers unique imaging opportunities with this respect.

### *4.3 Future Engineering Improvements*

An upgrade considered is the direct, active cooling of the coils, which would reduce the cooling times and potentially allow for continuous operation. Another modification considered is the replacement of the stainless steel central column with a central solenoid. This would allow for higher plasma currents, should the EC-driven currents not be sufficient.

## ACKNOWLEDGMENTS

Useful discussions are acknowledged with W. Reiersen (ORNL), who motivated and provided initial design parameters for the ITER-sized tilted coil configuration. Technical assistance by J. Andrello, P. Byrne, C. Caliri, A. Febre and T.M. Roberts is thankfully acknowledged. Finally, we thank R. Woodward for the help in structural calculations and B.Y. Israeli, A. Nielsen and N. Rivera for the fruitful discussions.

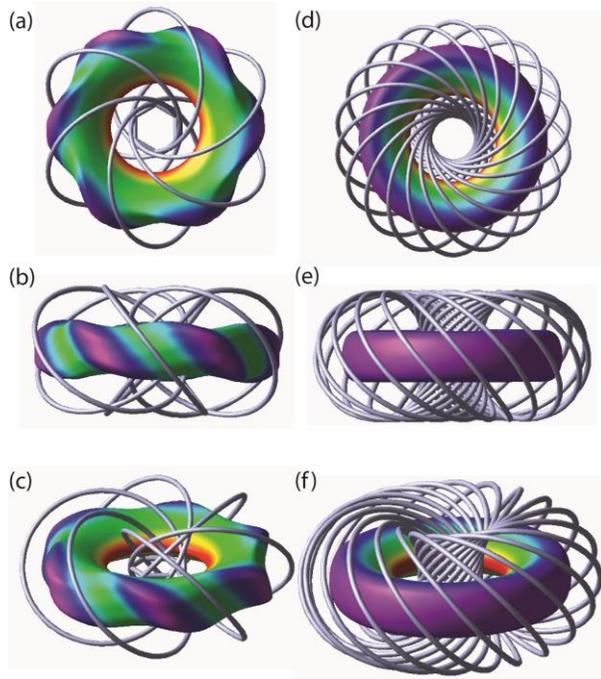

Fig. 1. Top, side and bird's-eye views of sets of 6 or 18 tilted and interlinked Toroidal Field coils. A finite plasma current and the addition of Vertical and Quadrupole Field coils (not shown) leads to three-dimensional equilibria [8] that were calculated with the VMEC code [10]. Shown are the last closed flux surfaces of such equilibria. The color contours denote the intensity of the magnetic field.

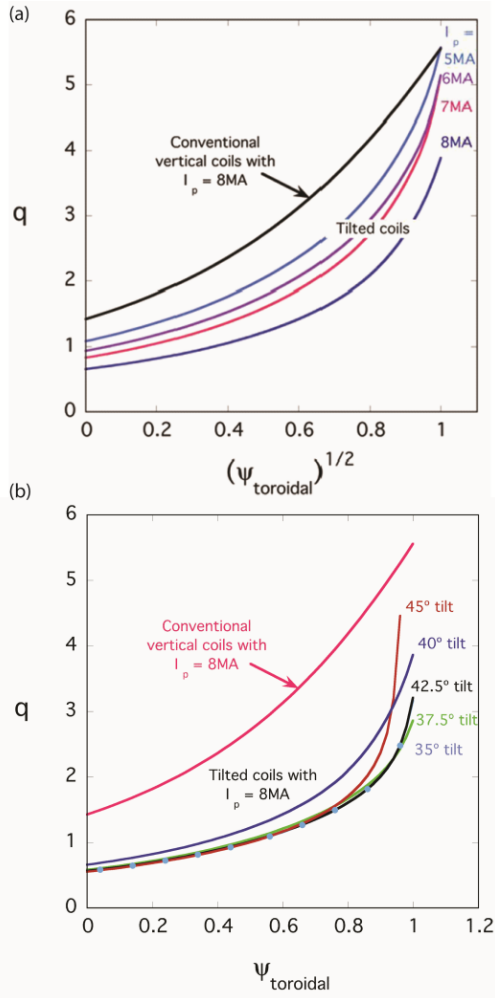

Fig. 2. Profile of safety factor $q$ in a toroidal device equipped with 18 tilted coils as a function of a flux-surface label (a) various plasma currents, for 40° tilt and (b) various tilt angles, for $I_p$=8 MA. Here $\Psi_{toroidal}$ is the normalized toroidal flux.

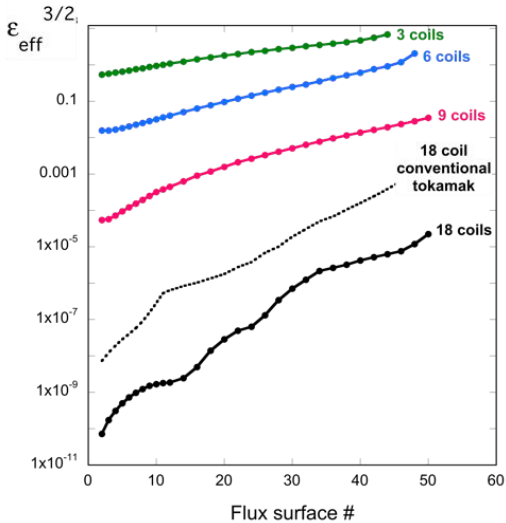

Fig. 3. Comparison of effective ripple, $\varepsilon_{eff}$, for tokamak-stellarator hybrids with various numbers of coils and a conventional 18-coil tokamak.

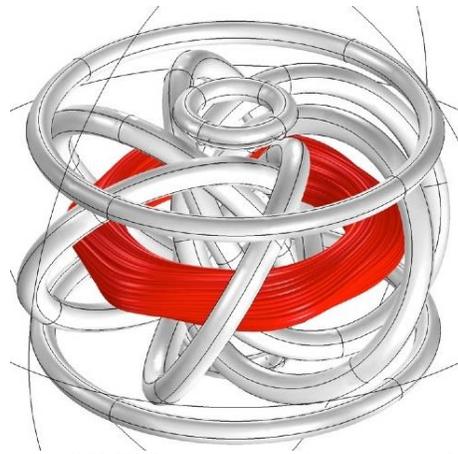

Fig. 4. Field lines on last closed flux surface traced by COMSOL software for plasma current $I_P$=2.5kA.

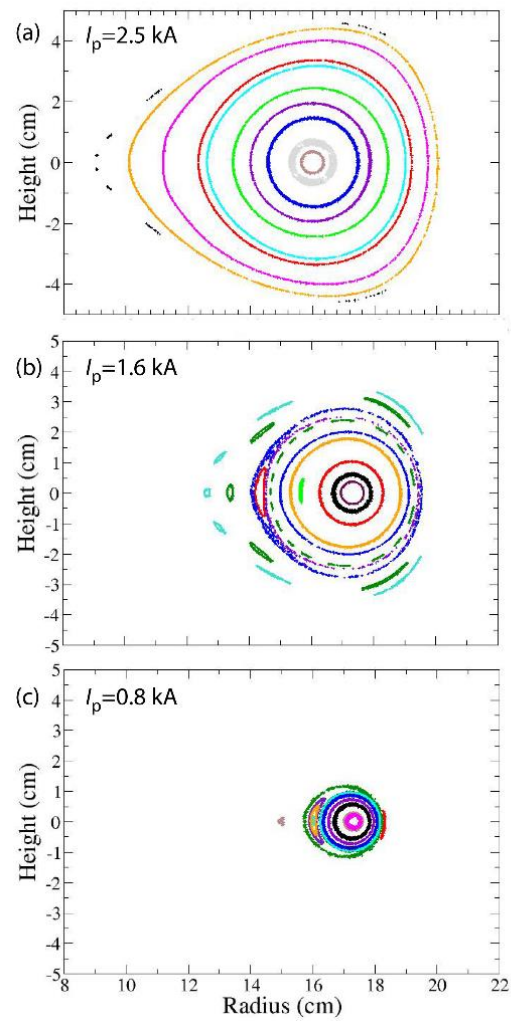

Fig. 5. Poincare plots showing flux surfaces for fixed coil-currents ($I_{TF}$=5.2 kA, $I_{VF}$=4.1 kA and $I_{QF}$=4.2 kA, as per Table 1) and for plasma-currents of, respectively, (a) $I_p$=2.5 kA, (b) 1.6 kA and (c) 0.8 kA.

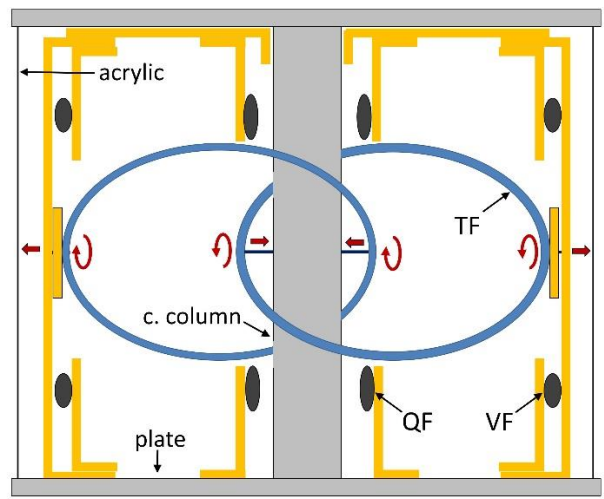

Fig. 6. Cross sectional schematics of stainless steel plates and central column (grey), acrylic vessel (black), Quadrupole Field (QF) and Vertical Field (VF) coils (also in black) and their support structures (orange). Also shown are two Toroidal Field (TF) coils (blue), but in side-view rather than cross-section. TF coils are shown at their innermost radial position. The coils can be moved outward by sliding their supports on the central column inward, and their outer support structures outward (red horizontal arrows). The tilt-angles of the TF coils can also be adjusted (red curved arrows).

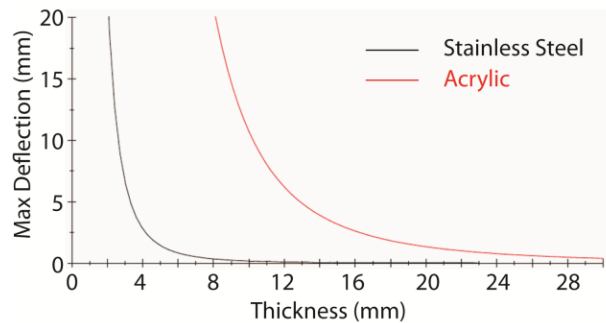

Fig. 7. Comparison of maximum deflection (mm) of the end caps as a function of their thickness.

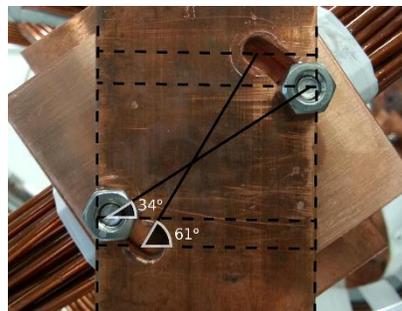

Fig. 8. Coil support bracket. Curvilinear slots allow to continuously adjust the coil tilt between 34 and 61° relative to the midplane.

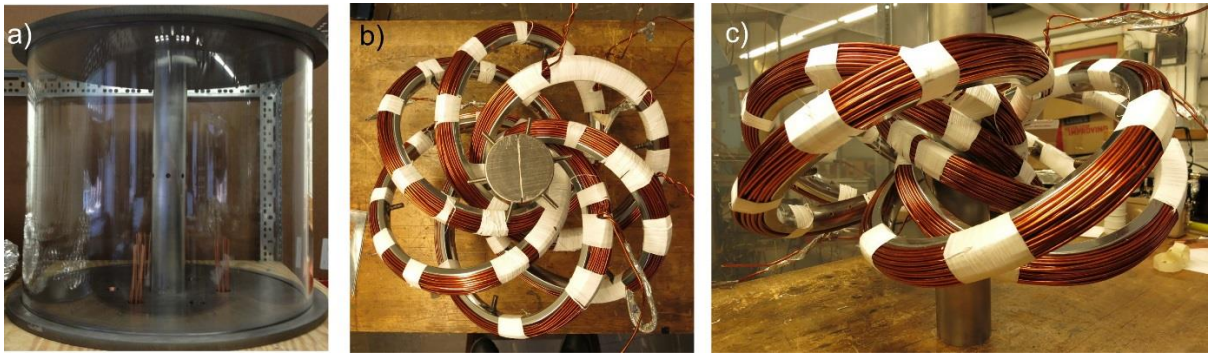

Fig. 9 (a) Acrylic chamber, central column and top and bottom plates assembled for vacuum test. (b) Top view and (c) side view of the tilted interlinked coils mounted on the central column before final assembly.

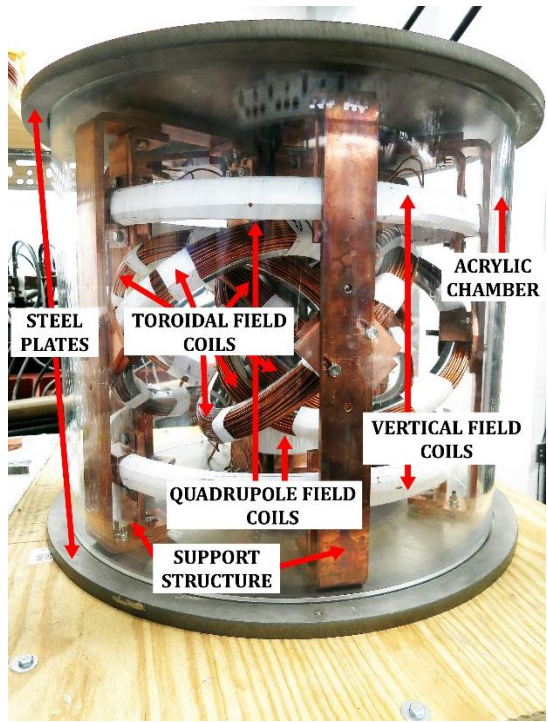

Fig.10 Final assembly of Proto-CIRCUS.

TABLE 1. Radial and vertical locations, $R$ and $Z$, of the centers of the coils, along with the coil radii $R_{coil}$, typical coil currents $I$ and other electrical specifications.

| Coil | $R$ (cm) | $Z$ (cm) | $R_{coil}$ (cm) | $I$ (kA) | Coils per series | No. Turns | $R$ (m$\Omega$) | $V$ (V) |
|---|---|---|---|---|---|---|---|---|
| TF | 16 | 0 | 10.7 | 5.2 | 2 | 69 | 297 | 22.3 |
| VF | 25.25 | 15.4 | 0 | 4.1 | 1 | 54 | 177 | 13.3 |
| QF | 7.25 | 15 | 0 | 4.2 | 2 | 56 | 95 | 7.1 |